\documentclass[12pt,aps]{revtex4}
\usepackage{psfig}

\begin{document}

\title{Boxed Plane Partitions as an Exactly Solvable Boson Model}

\author{N.M. Bogoliubov}

\affiliation{ St.~Petersburg Department of V.A. Steklov
Mathematical Institute, 27, Fontanka, St.~Petersburg 191023,
Russia}

\begin{abstract}
Plane partitions naturally appear in many problems of statistical
physics and quantum field theory, for instance, in the theory of
faceted crystals and of topological strings on Calabi-Yau
threefolds. In this paper a connection is made between the exactly
solvable model with the boson dynamical variables  and a problem
of enumeration of boxed plane partitions - three dimensional Young
diagrams placed into a box of a finite size. The correlation
functions of the boson model may be considered as the generating
functionals of the Young diagrams with the fixed heights of  its
certain columns. The evaluation of the correlation functions is
based on the Yang-Baxter algebra. The analytical answers are
obtained in terms of determinants and they can also be expressed
through the Schur functions.
\end{abstract}

\maketitle

\section{Introduction}
The theory of plane partitions is a classical chapter in
combinatorics \cite{Andrews}. Statistics of plane
partitions with respect to natural probabilistic measures was
studied in \cite{ver}, \cite{verker}. Locally, plane partitions
are equivalent to a tiling of a plane by rhombi (or lozenges).
Among many results obtained in this direction we should mention
the paper \cite{Cohnl} where the phenomenon of the "arctic circle"
was described for boxed plane partitions. In the paper \cite{Cohn}
the variational principle applicable to a variety of such problems
was developed. Correlation functions for random plane partitions
were studied in \cite{Cohn}, \cite{or}. For boxed plane partitions
distributed uniformly they were computed in the bulk of the limit
shape (inside of the arctic circle) in \cite{Cohn} , and in
\cite{or} for unrestricted plane partitions distributed as
$q^{|\pi|}$, where $|\pi|$ is a volume of a partition.

Plane partitions naturally appear in many problems of statistical
physics, for instance, in the theory of faceted crystals \cite{fs}
and of direct percolation \cite{rd}. Quite recently it was argued
that there was a connection between topological strings on
Calabi-Yau threefolds and crystal melting \cite{orv}.

In this paper we demonstrate  the connection of a certain
integrable boson type model and boxed plane partitions - the plane
partitions placed into a box of a finite size. The natural
dynamical variables in this boson model are the $q-$bosons. The
algebra of $q$-bosons \cite{dam} appear naturally within the
quantum algebra formalism \cite{frt}.

Boxed plane partitions are related to the special case of
$q-$bosons, namely to a limit when the deformation parameter $q$
tends to zero what corresponds to an infinite value of coupling
constant of the $q-$boson model. In this special limit $q-$bosons
are known as exponential phase operators of quantum non-linear
optics. Notice that this is also the famous crystal limit for
quantum groups \cite{kash}, \cite{lus}.

Our analysis is  based on the Quantum Inverse Scattering Method
(QISM) and on the algebraic approach to the calculation of the
correlation functions developed within this method \cite{fad},
\cite{kbi}. We shall show that the scalar product of the state
vectors of the phase model \cite{bbt}, \cite{bn}, \cite{bik} is
related to MacMahon enumeration formula for boxed plane
partitions. There is a well established connection between the
theory of plane partitions and the theory of random processes.
From that point of view the scalar product of a phase model is a
generator of a point fermion-like random field \cite {or}. The
systematic application of the QISM allows to calculate different
correlation functions appearing in the theory of the boxed plane
partitions.

\section{$q-$bosons}

The $q-$boson algebra is defined by three independent operators $%
B,B^{\dagger }$ and $N$ satisfying commutation relations
\begin{eqnarray}
\lbrack B,B^{\dagger }] &=&q^{2N},\,\,  \nonumber \\
\lbrack N,B^{\dagger }] &=&B^{\dagger },\,\,[N,B]=-B,  \label{qb}
\end{eqnarray}
and $q$, a $c$-number, is taken to be $q=e^{-\gamma }.$ We consider real $%
\gamma >0.$ The $q-$boson algebra (\ref{qb}) has the
representation in the Fock space formed from the $q$-boson
normalized states $|n\rangle $
\begin{equation}
B^{\dagger }|n\rangle =[n+1]^{\frac 12}|n+1\rangle
\,,\,\,\,\,B|n\rangle =[n]^{\frac 12}|n-1\rangle ,  \label{frep}
\end{equation}
where the ''box'' is
\begin{equation}
\lbrack n]=\frac{1-q^{2n}}{1-q^2}.  \label{box}
\end{equation}
The integer numbers $n>0$ are called occupation numbers or the
number of particles in a state $|n\rangle $:
\begin{equation}
N|n\rangle =n|n\rangle .  \label{qstate}
\end{equation}

If $q=1$ ($\gamma =0$), the $q-$bosons become ordinary bosons,
\begin{equation}
B\rightarrow b,B^{\dagger }\rightarrow b^{\dagger},N=b^{\dagger }b;\,\,\,[b,b^{\dagger }] =1.
\label{bos}
\end{equation}

In the limit $q\rightarrow 0\,(\gamma \rightarrow \infty )$ the operators $%
B,B^{\dagger }$ transform into the operators $\phi ,\phi ^{\dagger
}$ defined by the commutation relations
\begin{equation}
[N,\phi ]=-\phi \,,\,[N,\phi ^{\dagger }]=\phi ^{\dagger
},\,\,[\phi ,\phi ^{\dagger }]=\pi \,  \label{phalg}
\end{equation}
in which $\pi $ is the vacuum projector $\pi =|0\rangle \langle
0|.$ The Fock states $|n\rangle $ can be created from the vacuum
state $|0\rangle $ by operating by the phase operators,
\begin{equation}
|n\rangle =(\phi ^{\dagger })^n|0\rangle ,\,\ N|n\rangle =n|n\rangle , \label{phopa}
\end{equation}
and
\begin{equation}
\phi ^{\dagger }|n\rangle =|n+1\rangle ,\,\ \phi |n\rangle =|n-1\rangle ,\,\ \phi |0\rangle =0.  \label{phop}
\end{equation}
One may verify that $\phi $ and $\phi ^{\dagger }$ can be
expressed in terms of the Fock states, $|n\rangle ,$ as
\[
\phi =\sum_{n=0}^\infty |n\rangle \langle n+1|,\,\phi ^{\dagger
}=\sum_{n=0}^\infty |n+1\rangle \langle n|.
\]
The introduced operator $\phi $ is ''one-sided unitary'' or an
isometric, although
\[
\phi \phi ^{\dagger }=1,
\]
one has
\[
\phi ^{\dagger }\phi =1-|0\rangle \langle 0|.
\]
The operators (\ref{phalg}) may be expressed in terms of ordinary bosons (%
\ref{bos}):
\[
\phi =(b^{\dagger }b+1)^{-\frac 12}b,\,\phi ^{\dagger }=b^{\dagger
}(b^{\dagger }b+1)^{-\frac 12}.
\]

\section{Integrable phase model}

The phase model is a special limit of the integrable $q$-boson
model \cite{bbt}, \cite {kul}. It is defined by the $L-$operator
\cite{bik}, \cite{bn}:
\begin{equation}
L_n(u)\equiv \left(
\begin{array}{cc}
a_n(u) & b_n(u) \\
c_n(u) & d_n(u)
\end{array}
\right) =\left(
\begin{array}{cc}
u^{-1} & \phi _n^{\dagger } \\
\phi _n & u
\end{array}
\right) ,  \label{lop}
\end{equation}
where parameter $u\in {\cal C}$, and $\phi _n,\phi _n^{\dagger }$
are the operators (\ref{phalg}) satisfying commutation relations
\begin{equation}
\lbrack N_i,\phi _j]=-\phi _i\delta _{ij}\,,\,[N_i,\phi
_j^{\dagger }]=\phi _i^{\dagger }\delta _{ij}\,,\,\,[\phi _i,\phi
_j^{\dagger }]=\pi _i\delta _{ij}\,.  \label{phaseal}
\end{equation}
On the local Fock vectors
\begin{eqnarray}
\phi _j|0\rangle_j&=&0,\nonumber \\
\phi _j|n_j\rangle _j &=&|n_j-1\rangle _j,
\phi_j^{\dagger }|n_j\rangle _j=|n_j+1\rangle _j, \label{phact}\\
N_j|n_j\rangle _j &=&n_j|n_j\rangle _j.  \nonumber
\end{eqnarray}
The operator valued matrix (\ref{lop}) satisfies the intertwining
relation
\begin{equation}
R(u,v)\left( L_n(u)\otimes L_n(v)\right) =\left( L_n(v)\otimes
L_n(u)\right) R(u,v)  \label{rll}
\end{equation}
in which $R(u,v)$ is the $4\times 4$ matrix with the non-zero
elements equal to
\begin{eqnarray}
R_{11}(u,v) &=&R_{44}(u,v)=f(v,u),  \nonumber \\
R_{22}(u,v) &=&R_{33}(u,v)=g(v,u),  \label{r} \\
R_{23}(u,v) &=&1,  \nonumber
\end{eqnarray}
with
\begin{equation}
f(v,u)=\frac{u^2}{u^2-v^2},\,\,\,\,\,g(v,u)=\frac{uv}{u^2-v^2}.
\label{fg}
\end{equation}
Symbol $\otimes $ denotes the tensor product of matrices: $\left(
A\otimes B\right) _{ij,kl}=A_{ij}B_{kl}$. The $R$-matrix (\ref{r})
satisfies the Yang-Baxter equation
\begin{eqnarray}
&&\left( I\otimes R(u,v)\right) \left( R(u,w)\otimes I\right)
\left(
I\otimes R(v,w)\right)  \nonumber \\
&=&\left( R(v,w)\otimes I\right) \left( I\otimes R(u,w)\right)
\left( R(u,v)\otimes I\right) .  \label{yb}
\end{eqnarray}

The monodromy matrix is introduced as
\begin{equation}
T(u) =L_M(u)L_{M-1}(u)...L_0(u) =
\left(
\begin{array}{cc}
A(u) & B(u) \\
C(u) & D(u)
\end{array}
\right).  \label{mm}
\end{equation}
Matrix elements of this matrix act in the Fock space spanned on
the state vectors
\begin{equation}
|n\rangle =\prod_{j=0}^M|n_j\rangle _j=\prod_{j=0}^M(\phi
_j^{\dagger})^{n_j}|0\rangle ,\,\,\sum n_j=n, \label{stv}
\end{equation}
where
\begin{equation}
|0\rangle =\prod_{j=0}^M|0\rangle _j\,  \label{vac}
\end{equation}
is the vacuum vector (generating state), and $_k\langle n_m|n_i\rangle _j =
\delta _{im}\delta _{kj}$.

The commutation relations of the matrix elements of the monodromy
matrix are given by the $R-$matrix (\ref{r})
\begin{equation}
R(u,v)\left( T(u)\otimes T(v)\right) =\left( T(v)\otimes
T(u)\right) R(u,v). \label{rtt}
\end{equation}
The most important relations are
\begin{eqnarray}
C(u)B(v) &=&g(u,v)\left\{ A(u)D(v)-A(v)D(u)\right\},  \label{cb} \\
C(u)A(v) &=&f(v,u)A(v)C(u)+g(u,v)A(u)C(v),  \label{ab} \\
D(u)B(v) &=&f(v,u)B(v)D(u)+g(u,v)B(u)D(v),  \label{db} \\
\lbrack B(u),B(v)] &=&[C(u),C(v)]=0.  \label{bbcc}
\end{eqnarray}
The relation (\ref{rtt}) means that the transfer matrix
$\tau (u)=trT(u)=A(u)+D(u)$ is the generating function
of the integrals of motion: $[\tau (u),\tau (v)]=0$ for all
$u,v\in {\cal C}$.

The $L-$operator (\ref{lop}) satisfies the relation
\begin{equation}
e^{\zeta N_n}L_n(u)e^{\frac 12\zeta \sigma ^z}=e^{\frac 12\zeta
\sigma ^z}L_n(u)e^{\zeta N_n};\sigma ^z=\left(
\begin{array}{cc}
1 & 0 \\
0 & -1
\end{array}
\right),  \label{ln}
\end{equation}
where $\zeta \in {\cal C},$ and $N_n$ is the number operator
(\ref{phact}). From this equation and the definition of the
monodromy matrix (\ref{mm}) it follows that
\begin{equation}
e^{\zeta \hat N}T(u)e^{\frac 12\zeta \sigma ^z}=e^{\frac 12\zeta
\sigma ^z}T(u)e^{\zeta \hat N},  \label{mn}
\end{equation}
where
\begin{equation}
\hat N=\sum_{n=0}^MN_n  \label{tn}
\end{equation}
is a total number operator, and $\sigma ^z$ is the Pauli matrix
(\ref{ln}). The equation (\ref{mn}) is equivalent to
\begin{eqnarray}
\hat NB(u) &=&B(u)\left( \hat N+1\right) ,  \label{nbc} \\
\hat NC(u) &=&C(u)\left( \hat N-1\right) .  \nonumber
\end{eqnarray}
It means that the operator $B(u)$ is a creation operator, while
$C(u)$ is annihilation one.

The generating vector $|0\rangle $ (\ref{vac}) is annihilated by
$C(u)$
operator and is the eigenvector of $A(u)$ and $D(u):$%
\begin{eqnarray}
C(u)|0\rangle &=&0,\label{aadd}  \\
A(u)|0\rangle &=&a(u)|0\rangle ,\,D(u)|0\rangle =d(u)|0\rangle , \nonumber
\end{eqnarray}
with the eigenvalues $a(u)=u^{-(M+1)}$ and $d(u)=u^{M+1}$ respectively. The $%
N-$particle vectors are taken to be of the form
\begin{equation}
|\Psi _N(u_1,u_2,...,u_N)\rangle =\prod_{j=1}^NB(u_j)|0\rangle ,
\label{bstv}
\end{equation}
and
\[
\hat N|\Psi _N(u_1,u_2,...,u_N)\rangle =N|\Psi
_N(u_1,u_2,...,u_N)\rangle .
\]
The vacuum state (\ref{vac}) is similar to the highest-weight
vector in the theory of representations of Lie algebras. The state
conjugated to (\ref {bstv}) is
\begin{equation}
\langle \Psi _N(u_1,u_2,...,u_N)|=\langle 0|\prod_{j=1}^NC(u_j).
\label{cstv}
\end{equation}
The dual vacuum $\langle 0|=|0\rangle ^{\dagger },$ $\langle
0|0\rangle =1.$ It is easy to verify that $\langle 0|B(u)=0$ and
$\langle 0|A(u)=a(u)\langle 0|,$ $\langle 0|D(u)=d(u)\langle 0|.$

One can visualize matrix elements of the $L-$ operator as a vertex with the attached
arrows (see FIG. 1). The matrix element $b_n(u)=\phi _n^{\dagger }$
corresponds to a vertex (b), $c_n(u)=\phi _n$ corresponds to a vertex (c),
$a_n(u)=u^{-1}$ to a vertex (a) and $d_n(u)=u$ to a vertex (d) respectively.
\begin{figure}
\centerline{\psfig{file=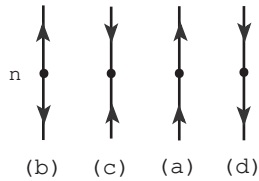}} \caption{Vertex representation
of the matrix elements of the $L$-operator.}
\end{figure}

Matrix elements of the monodromy matrix (\ref{mm}) are expressed then as sums over all
possible configurations of arrows with different boundary conditions on a one-dimensional
lattice with $M+1$ sites (see FIG. 2). Namely, operator $B(u)$ corresponds to the
boundary conditions when arrows on the top and bottom of the lattice are
pointing outward (configuration (B)). Operator $C(u)$ corresponds to the boundary
conditions when arrows on the top and bottom of the lattice are pointing outward
(configuration (C)). Operators $A(u)$ and $D(u)$ correspond to the
boundary conditions when arrows on the top and bottom of the lattice
are pointing up and down respectively (configurations (A) and (D)):
\begin{figure}[h]
\centerline{\psfig{file=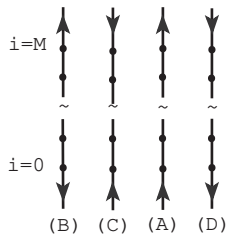}} \caption{Matrix elements of
the monodromy matrix $T(u)$.}
\end{figure}

For example an operator $B(u)$ on the lattice of three sites $(M=2)$ is: $%
B(u)=\phi _0^{\dagger }u^{-2}+\phi _0^{\dagger }\phi _1\phi
_2^{\dagger }+u^2\phi _2^{\dagger }+\phi _1^{\dagger },$ and may
be represented in a form (see FIG. 3).
\begin{figure}[h] \centerline{\psfig{file=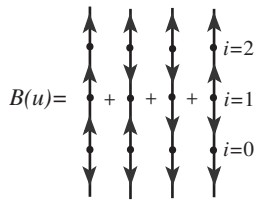}}
\caption{Vertex representation of the operator $B(u)$.}
\end{figure}

\section{Scalar products and plane partitions}

Recall that a partition $\lambda =(\lambda _1,\lambda
_2,...,\lambda _N)$ is a non-increasing sequence of non-negative
integers $\lambda _1\geq \lambda _2\geq ...\geq \lambda _N$
called the parts of $\lambda $. The sum of the parts of $\lambda $
is denoted by $|\lambda |$.

A plane partition is an array $\pi _{ij}$ of non-negative
integers that are non-increasing as functions of both $i$ and $j$
$(i,j=1,2,...).$ The integers $\pi _{ij}$ are called the parts of
the plane partition, and $|\pi |=\sum \pi _{ij}$ is its volume.
The plane partitions are often interpreted as stacks of cubes
(three-dimensional Young diagrams). The height of a stack with
coordinates $ij$ is $\pi _{ij}.$ If we have $i\leq r,j\leq s$ and
$\pi _{ij}\leq t$ for all cubes of the plane partition, it is said
that the plane partition is contained in a box with side lengths
$r,s,t.$  The symmetric plane partition is the plane partition for
which $\pi _{ij}=\pi _{ji}$. In (FIG. 4) the diagram corresponding
to the boxed plane partition $4\times 4\times 5$
\begin{equation}
\pi =\left(
\begin{array}{cccc}
5 & 3 & 3 & 2 \\
5 & 2 & 1 & 1 \\
4 & 1 & 1 & 0 \\
3 & 1 & 1 & 0
\end{array}
\right)  \label{pi}
\end{equation}
is represented.
\begin{figure}[h]
\centerline{\psfig{file=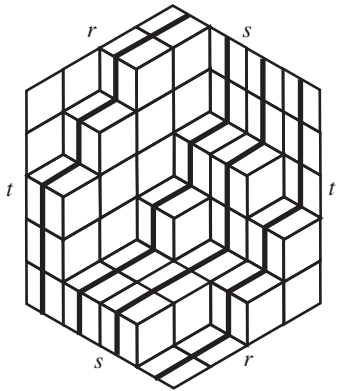}} \caption{A plane partition with
a gradient lines.}
\end{figure}

A plane partition in a $r\times s\times t$ box is equivalent to a
lozenge tiling of an $(r,s,t)$-semiregular hexagon. The term
lozenge refers to a unit rhombi with angles of $\frac \pi 3$ and
$\frac{2\pi }3.$
\begin{figure}[h]
\centerline{\psfig{file=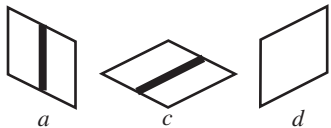}} \caption{Three types of
lozenges.}
\end{figure}
To each plane partition we can put into correspondence a $q-$weight $q^{|\pi |}$. Notice that these $q$ are not in
any way related to the $q$-deformation parameter of Section II. The sum of $q-$weights of all plane partitions
contained in a box $r\times s\times t$ is known as $q-$enumeration of plane partitions or the partition function of
Young diagrams
\begin{equation}
Z_q(r,s,t)=\sum_{p.p}q^{|\pi |}=\prod_{i=1}^r\prod_{j=1}^s\prod_{k=1}^t\frac{%
1-q^{i+j+k-1}}{1-q^{i+j+k-2}}=\prod_{j=1}^r\prod_{k=1}^s\frac{1-q^{t+j+k-1}}{%
1-q^{k+j-1}}.  \label{macmah}
\end{equation}
This formula is MacMahon generation function for the boxed plane
partitions \cite{bres}, \cite{macd}. In the $q\rightarrow 1$ limit
this formula gives the number of the plane partitions containing
in a box $r\times s\times t$.

Let us consider the scalar product of the state vectors (\ref{bstv}) and (%
\ref{cstv}):
\begin{equation}
S(N,M|\{v\},\{u\})=\langle
0|\prod_{j=1}^NC(v_j)\prod_{j=1}^NB(u_j)|0\rangle ,  \label{scpr}
\end{equation}
where $\{u\}$ and $\{v\}$ are the sets of independent parameters.
We shall show that after the parametrization
\begin{equation}
u_j=q^{\frac{(j-1)}2},\,\,v_j=q^{-\frac j2},\,  \label{hlip}
\end{equation}
the scalar product (\ref{scpr}) is the generating function
(\ref{macmah}) of the plane partitions containing in a box
$N\times N\times M$:
\begin{equation}
q^{\frac{N^2M}2}\langle 0|\prod_{j=-1}^{-N}C(q^{\frac j2})\prod_{j=1}^NB(q^{%
\frac{j-1}2})|0\rangle =Z_q(N,N,M).  \label{sa}
\end{equation}

To make a connection of the scalar product (\ref{scpr}) with the problem of enumeration
of boxed plane partitions we shall use a graphic representation of its elements introduced
in the previous Section. Consider a two-dimensional square lattice with $2N\times (M+1)$ sites.
First $N$ vertical rows of the lattice are associated with the operators $C(v_j)$ and the
last $N$ vertical rows
with operators $B(u_j)$ (see FIG. 1 and FIG. 2). The horizontal rows of the lattice are associated with the local
Fock spaces, $i$-th raw with the $i$-th space respectively. Each horizontal edge of a lattice is
labelled then with an occupation number $n_j$ of a correspondent Fock vector $%
|n_j\rangle _i.$ The scalar product (\ref{scpr}) is equal to the
sum over all allowed configurations on a square lattice with the
arrows on the first $N$ vertical rows pointing inwards, on the last
$N$ ones pointing outwards; on the right and on the left
boundaries all occupation numbers are zeros.

These configurations may be represented in terms of
the lattice paths. The allowed configurations are then the number
of possible
non-crossing $N$ paths starting from the down left $N$ lattice edges $%
(-N;0),(-N+1;0),...,(-1;0)$ and ending at the $N$ top right ones $%
(1;M),(2;M),...,(N;M)$. The $m-$th path is running from $(-N+m-1;0)$ to $%
(m;M)$, $1\leq m\leq N$. In the vertical direction paths follow
the arrows, and only one path is allowed on a vertical lattice
edge but any number of paths can share the horizontal ones. The
number of paths sharing a horizontal edges is equal to the
corresponding occupation number of the edge. The length of the
paths is $(N+M).$ One of the possible configurations is
represented in (FIG. 6).
\begin{figure}[h]
\centerline{\psfig{file=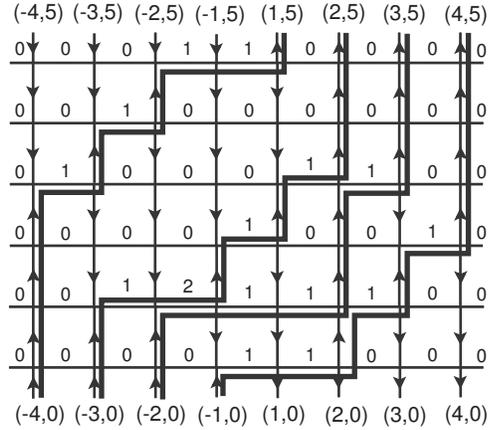}} \caption{A typical
configuration of admissible lattice paths.}
\end{figure}

The cells of the lattice under the $m-$th path may be considered
as diagram of corresponding partition and may be thought of as the
$m-$th column in the array $(\pi _{i,j})$. The configuration of
the paths in (FIG. 6) corresponds to the plane partition in (FIG.
4) and respectively to the array (\ref{pi}).

On the other hand we can associate the vertical and horizontal
edges carrying paths with lozenges. Lozenge ($a$) in (FIG. 5)
corresponds to a vertical line of the path, while lozenge ($c$) to
a horizontal one. Lattice edges without the paths correspond to a
lozenge ($d$). The lozenge tiling is simply the projection of
three-dimensional Young diagram with gradient lines. This
establishes the mapping of the configurations generated by the
scalar product (\ref{scpr}) on the plane partitions.

Consider the scalar product (\ref{scpr}). Due to commutation
relations (\ref {bbcc}) it is a symmetric function of $N$
variables $v_j$ and also a symmetric function of $N$ variables
$u_j.$ It is easy to verify that the number of operators $C(u)$
should be equal to the number of operators $B(u),$ otherwise the
scalar product is equal to zero. The scalar product is evaluated
by means of commutation relations (\ref{cb})-(\ref{db}). In the
simplest case ($N=1$) the scalar product is equal to
\begin{eqnarray*}
S(1,M|v,u) &=&\langle 0|C(v)B(u)|0\rangle \\
\ &=&g(v,u)\left\{ a(v)d(u)-a(u)d(v)\right\} \\
&=&\left\{ \left( \frac uv\right) ^{M+1}-\left( \frac uv\right)
^{-M-1}\right\} \times \frac 1{\frac uv-\left( \frac uv\right)
^{-1}},
\end{eqnarray*}
where $g(u,v)$ is the element of $R$-matrix (\ref{fg}) and $a(u)$
and $d(u)$ are the eigenvalues of $A(u)$ and $D(u)$ (\ref{aadd}).

For the arbitrary $N$ one may get \cite{kbi}, \cite{bik}:
\begin{eqnarray}
S(N,M|\{v\},\{u\}) &=&\langle
0|\prod_{j=1}^NC(v_j)\prod_{j=1}^NB(u_j)|0\rangle  \nonumber \\
\ &=&\left\{ \prod_{j>k}\left(
\frac{v_k}{v_j}-\frac{v_j}{v_k}\right) ^{-1}\prod_{l>m}\left(
\frac{u_l}{u_m}-\frac{u_m}{u_l}\right) ^{-1}\right\} \det H,
\label{sdet}
\end{eqnarray}
where the matrix elements of $N\times N$ matrix $H$ are equal to
\begin{equation}
H_{jk}=\left\{ \left( \frac{u_k}{v_j}\right) ^{M+N}-\left( \frac{u_k}{v_j}%
\right) ^{-M-N}\right\} \times \frac 1{\frac{u_k}{v_j}-\left( \frac{u_k}{v_j}%
\right) ^{-1}}.  \label{t}
\end{equation}
The parametrization (\ref{hlip}) transforms the scalar product
into
\begin{equation}
S(N,M|\{q\})=(-1)^{\frac{N(N-1)}2}\left\{ \prod_{j>k}\left( q^{\frac{j-k}2%
}-q^{-\frac{j-k}2}\right) ^{-2}\right\} \det {\cal H},  \label{se}
\end{equation}
where
\begin{equation}
{\cal H}_{jk}=\frac{s^{\frac{k+j-1}2}-s^{-\frac{k+j-1}2}}{q^{\frac{k+j-1}2%
}-q^{-\frac{k+j-1}2}},  \label{hme}
\end{equation}
with $s=q^{M+N}.$ The determinant of the matrix ${\cal H}$ was
considered in \cite{kup} in connection with the alternating sign
matrices enumeration problem and is equal to
\begin{equation}
\det {\cal H}=(-1)^{\frac{N(N-1)}2}\left\{ \prod_{j>k}\left( q^{\frac{j-k}2 }-q^{-\frac{j-k}2}\right) ^2\right\}
\prod_{1\leq j,k\leq N}\frac{s^{\frac 12}q^{\frac{j-k}2}-s^{-\frac 12}q^{-\frac{j-k}2}}
{q^{\frac{k+j-1}2}-q^{-\frac{k+j-1}2}}.  \label{kup}
\end{equation}
Therefore,
\begin{eqnarray}
S(N,M|\{q\}) &=&q^{-\frac{N^2M}2}\prod_{1\leq j,k\leq N}\frac{1-q^{N+M+j-k}}{1-q^{k+j-1}}  \nonumber \\
\ &=&q^{-\frac{N^2M}2}\prod_{1\leq j,k\leq N}\frac{1-q^{M-1+j+k}}{1-q^{k+j-1}}.  \label{hse}
\end{eqnarray}
Finally, we have obtained the equality (\ref{sa}) for the scalar
product:
\begin{equation}
\langle 0|C(q^{-\frac N2})...C(q^{-1})C(q^{-\frac 12})B(1)B(q^{\frac 12})...B(q^{\frac{N-1}2})|0\rangle
=q^{-\frac{N^2M}2}Z_q(N,N,M). \label{sppp}
\end{equation}

\section{Coordinate form of state vectors and Schur functions}

Using the explicit form of the operators $B(\lambda )$ we may rewrite the $%
N- $particle state vector (\ref{bstv}) in the ''coordinate'' form
\begin{equation}
|\Psi _N(u_1,u_2,...,u_N)\rangle =\prod_{k=1}^NB(u_k)|0\rangle =\sum_{%
{0\leq n_0,n_1,...,n_M\leq N \atop n_0+n_1+...+n_M=N}
}f_{\{n\}}(u_1,u_2,...,u_N)\prod_{j=0}^M|n_j\rangle _j,
\label{cbawf}
\end{equation}
with the function $f$ equal to
\begin{equation}
f_{\{n%
\}}(u_1,u_2,...,u_N)=f_{(n_{j_1},n_{j_2,}...,n_{j_M})}(u_1,u_2,...,u_N)=%
\sum_{{\cal B}}u_1^{t_1^d-t_1^a}u_2^{t_2^d-t_2^a}\cdot ...\cdot
u_N^{t_N^d-t_N^a}.  \label{ampl}
\end{equation}
Here the sum is taken over all admissible $N$ paths with $n_{j_1}$
paths
starting from $(1;j_1)$, $n_{j_2}$ from $(1;j_2)$, and $n_{j_M}$ from $%
(1;j_M)$ respectively,
$j_1>j_2>...>j_M;n_{j_1}+...+n_{j_M}=N,\,n_{j_k}\neq
0 $. The power $t_k^d$ is equal to the number of the $d(u)$ vertices in the $%
k- $th vertical line of the grid, while $t_k^a$ is equal to the number of $%
a(u)$ vertices respectively.
\begin{figure}[h]
\centerline{\psfig{file=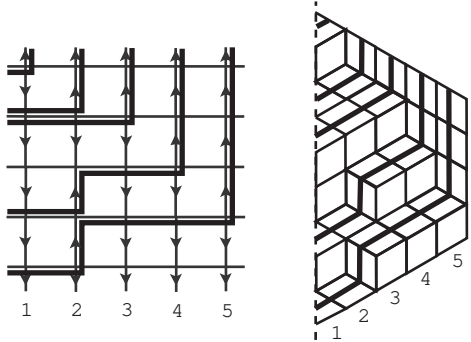}} \caption{Lattice paths
representation of a particular term of the 5-particle state
vector, and a correspondent part of a plane partition.}
\end{figure}

The following representation is valid
\begin{equation}
f_{\{n\}}(u_1,u_2,...,u_N)=\left( u_1u_2\cdot ...\cdot u_N\right)
^{-M}S_{\{\lambda \}}(u_1^2,u_2^2,...,u_N^2),  \label{cbafs}
\end{equation}
where $S_{\{\lambda \}}(u_1,u_2,...,u_N)$ is the Schur function
\begin{equation}
S_{(\lambda _1,...,\lambda _N)}(u_1,u_2,...,u_N)=\frac{\det \left(
u_j^{N-i+\lambda _i}\right) }{\prod_{1\leq i<j\leq N}(u_i-u_j)},
\label{schur}
\end{equation}
and there is one to one correspondence between the occupation
number configuration $\{n\}$ and partitions $\{\lambda \}$:

\begin{eqnarray}
\{n\} &=&(n_{j_1},n_{j_2,}...,n_{j_M})\Longleftrightarrow
\{\lambda
\}=(\lambda _1,...,\lambda _N)  \label{fpp} \\
\lambda _k &=&j_1;\,\,\,1\leq k\leq n_{j_1}  \nonumber \\
\lambda _k &=&j_2;\,\,\,n_{j_1}+1\leq k\leq n_{j_2}+n_{j_1}  \nonumber \\
&&\ \ \ \ \ \ \ ...  \nonumber \\
\lambda _k &=&j_N;\,\,\,n_{j_1}+...+n_{j_{M-1}}+1\leq k\leq
n_{j_1}+...+n_{j_M}=N.  \nonumber
\end{eqnarray}
The state conjugated to (\ref{cbawf}) is
\begin{equation}
\langle \Psi _N(u_1,u_2,...,u_N)|=\langle 0|\prod_{k=1}^NC(u_k)=\sum_{%
{0\leq n_0,n_1,...,n_M\leq N \atop n_0+n_1+...+n_M=N}
}f_{\{n\}}(u_1^{-1},u_2^{-2},...,u_N^{-2})\prod_{j=0}^M\langle
n_j|_j, \label{cbawfc}
\end{equation}
with
\begin{equation}
f_{\{n\}}(u_1^{-1},u_2^{-2},...,u_N^{-2})=\sum_{{\cal C}%
}u_1^{t_1^d-t_1^a}u_2^{t_2^d-t_2^a}\cdot ...\cdot
u_N^{t_N^d-t_N^a}, \label{amplc}
\end{equation}
where the sum is taken over all admissible $N$ paths with
$n_{j_1}$ paths ending at $(-1;j_1)$, $n_{j_2}$ ending at
$(-1;j_2)$, and $n_{j_M}$ ending
at $(1;j_M)$ respectively, $j_1>j_2>...>j_M;n_{j_1}+...+n_{j_M}=N,\,n_{j_k}%
\neq 0$.
\begin{figure}[h]
\centerline{\psfig{file=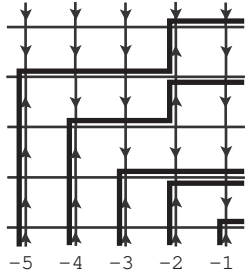}} \caption{Lattice paths
representation of a particular term of the conjugated state
vector.}
\end{figure}

Due to the orthogonality of the Fock states (\ref{stv}) the scalar
product is equal to
\begin{eqnarray}
S(N,M|\{v\},\{u\}) =\langle \Psi _N(v_1,v_2,...,v_N)|\Psi
_N(u_1,u_2,...,u_N)\rangle  \nonumber \\
=\sum_{{0\leq n_0,n_1,...,n_M\leq N \atop n_0+n_1+...+n_M=N}
}f_{\{n\}}(v_1^{-1},v_2^{-1},...,v_N^{-1})f_{\{n\}}(u_1,u_2,...,u_N)
\nonumber \\
=\sum_{{0\leq n_0,n_1,...,n_M\leq N \atop
n_0+n_1+...+n_M=N}}\sum_{{\cal
C}}v_1^{t_1^d-t_1^a}v_2^{t_2^d-t_2^a}\cdot ...\cdot
v_N^{t_N^d-t_N^a}\sum_{{\cal
B}}u_1^{t_1^d-t_1^a}u_2^{t_2^d-t_2^a}\cdot ...\cdot
u_N^{t_N^d-t_N^a}. \label{scu}
\end{eqnarray}
Taking into account the representation (\ref{cbafs}) we can express the scalar product in terms of Schur functions
\begin{equation}
S(N,M|\{v\},\{u\})=\left( \prod_{j=1}^N\frac{v_j^M}{u_j^M}\right) \sum_{\lambda \subseteq \{M^N\}}S_{\{\lambda
\}}(u_1^2,u_2^2,...,u_N^2)S_{\{\lambda \}}(v_1^{-2},v_2^{-2},...,v_N^{-2}), \label{scsch}
\end{equation}
where the sum is over all partitions, $\lambda $ , into at most
$N$ parts each of which is less than or equal to $M$. Comparing
this formula with (\ref {sdet}) we obtain the following
determinantal expression
\begin{eqnarray}
&&\ \ \sum_{\lambda \subseteq \{M^N\}}S_{\{\lambda
\}}(x_1^2,x_2^2,...,x_N^2)S_{\{\lambda \}}(y_1^2,y_2^2,...,y_N^2)
\label{scdet} \\
\ &=&\left( \prod_{j=1}^Nx_j^My_j^M\right) \left\{ \prod_{j>k}\frac{y_jy_k}{%
y_j^2-y_k^2}\prod_{l>m}\frac{x_lx_m}{x_l^2-x_m^2}\right\} \det H,
\nonumber
\end{eqnarray}
where the matrix $H$ is (\ref{t})
\[
H_{jk}=\left\{ \left( x_ky_j\right) ^{M+N}-\left( x_ky_j\right)
^{-M-N}\right\} \times \frac 1{x_ky_j-\left( x_ky_j\right) ^{-1}}.
\]

The volume $|\pi |$ of the plane partition $\pi$ in a box $(N\times N\times M)$ may be expressed as
\begin{equation}
2|\pi |=N^2M+\sum_{k=-1}^{-N}k\left( l_k^d-l_k^a\right)
+\sum_{j=1}^N(j-1)\left( l_j^d-l_j^a\right) ,  \label{volplp}
\end{equation}
where the first sum in this equality is over the columns going along the ''$%
s $'' side of hexagon while the second one is over the columns along the ''$%
r $'' side respectively, and $l_j^{a,d}$ is the number of lozenge of type $%
a,d$ (see FIG. 5) in the $j$-th column of the hexagon. It may be
checked that $l_j^d-l_j^a=t_j^d-t_j^a$. The substitution of the
parametrization (\ref {hlip}) into the equation (\ref{scu}) gives
for the scalar product
\begin{eqnarray*}
S(N,M|\{q\}) &=&\langle \Psi _N(q^{-\frac 12},q^{-1},...,q^{-\frac
N2})|\Psi
_N(1,q,...,q^{\frac{N-1}2})\rangle \\
\ &=&q^{-\frac{N^2M}2}\sum_{p.p}q^{|\pi |}.
\end{eqnarray*}
Together with (\ref{hse}) this equation provides us with the proof
of MacMahons enumeration formula for the boxed plane partitions
within the frames of Quantum Inverse Method.

From the equation (\ref{scsch}) we obtain the equality
\cite{kratt}:
\[
\sum_{\lambda \subseteq \{M^N\}}S_{\{\lambda
\}}(1,q,...,q^{N-1})S_{\{\lambda
\}}(q,q^2,...,q^N)=\sum_{p.p}q^{|\pi |}.
\]

The parts $\lambda _k$ of the partition $\lambda $ (\ref{fpp}) may be considered as the coordinates of the
particles (the coordinate $\lambda _k$ corresponds to the $k$-th particle). The state of the system is spanned by
the orthonormal basis $|\lambda \rangle =|\lambda _1,\lambda _2,...,\lambda _m\rangle :\,\,\langle \mu |\lambda
\rangle =\delta _{\{\mu \},\{\lambda \}}$. The $N$-particle state vector (\ref{cbawf}) is given then by the
equation
\begin{equation}
|\Psi _N(u_1,u_2,...,u_N)\rangle =\left( u_1u_2\cdot ...\cdot
u_N\right) ^{-M}\sum_{\lambda \subseteq \{M^N\}}S_{\{\lambda
\}}(u_1^2,u_2^2,...,u_N^2)|\lambda \rangle ,  \label{sw}
\end{equation}
and respectively
\begin{equation}
\prod_{j=1}^N\tilde B(u_j)|0\rangle =\sum_{\lambda \subseteq
\{M^N\}}S_{\{\lambda \}}(u_1^2,u_2^2,...,u_N^2)|\lambda \rangle ,
\label{sbw}
\end{equation}
where the sum in both formulas is taken over all partitions,
$\lambda $ , into at most $N$ parts each of which is less than or
equal to $M$. By the
construction operators $\tilde B(u)\equiv u^{-M}B(u)$, and $\tilde C%
(u)\equiv u^MC(u)$ possess the following properties
\begin{equation}
\tilde B(1)|\lambda \rangle  =\sum_{\mu \supset \lambda }|\mu
\rangle ,\,\,\,\,
\langle \lambda |\tilde C(1) =\sum_{\mu \supset \lambda }\langle
\mu |, \label{bcs}
\end{equation}
and
\begin{eqnarray}
\prod_{j=1}^N\tilde B(u_j)|\lambda \rangle  &=&\sum_{\mu \supset
\lambda
}S_{\{\mu /\lambda \}}(u_1^2,u_2^2,...,u_N^2)|\mu \rangle ,  \label{ssf} \\
\langle \lambda |\prod_{j=1}^N\tilde C(u_j) &=&\sum_{\mu \supset
\lambda }S_{\{\mu /\lambda
\}}(u_1^{-2},u_2^{-2},...,u_N^{-2})\langle \mu |, \nonumber
\end{eqnarray}
where $S_{\{\mu /\lambda \}}(x_1,x_2,...,x_N)$ is a skew Schur
function indexed by a pair of partitions $\mu $ and $\lambda $
such that $\lambda \subset \mu $ \cite{macd}. From this point of
view operators $\tilde B(u)$ and $\tilde C(u)$ may be considered
as the transition operators between the diagonals of a plane
partitions.

\section{Correlation functions}

Let us calculate the generating function of the plain partitions
contained in a box $N\times N\times M$ provided that the height
$\pi _{1N}$ of the stack of the cubes is fixed and equal to $m$.
\begin{figure}[t]
\centerline{\psfig{file=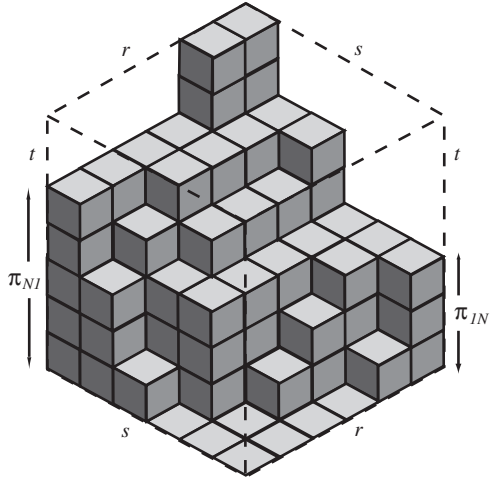}} \caption{A Young diagram
with fixed heights $\pi $ of the stack of cubes.}
\end{figure}
To find this function we have to consider a scalar product on a
lattice under the condition that the $N$-th lattice path enters the
$N$-th vertical line of the grid at the $m$-th row:
\begin{equation}
P_{1N}(m|\{v\},\{u\})=\langle
0|\prod_{j=1}^NC(v_j)\prod_{j=1}^{N-1}B(u_j)\phi _m^{\dagger
}|0\rangle . \label{mca}
\end{equation}
To calculate this scalar product we may use the following
decomposition of the operators $B(u)$ and $C(u)$:
\begin{eqnarray}
B(u)|0\rangle &=&u^{-M}\sum_{j=0}^M\phi _j^{\dagger
}u^{2j}|0\rangle ,
\label{brep} \\
\langle 0|C(u) &=&u^M\langle 0|\sum_{j=0}^M\phi _ju^{-2j}.
\label{crep}
\end{eqnarray}
The substitution of decomposition (\ref{brep}) into the scalar
product (\ref {scpr}) gives
\begin{equation}
S(N,M|\{v\},\{u\})=\sum_{m=0}^M(u_N)^{-M+2m}P_{1N}(m|\{v\},\{u\}).
\label{decsc}
\end{equation}
The determinant of the matrix $H$ in (\ref{sdet}) may be developed
by the last column. The comparison of the obtained decomposition
with (\ref{decsc}) leads to the equality
\begin{equation}
\frac{P_{1N}(m|\{v\},\{u\})}{S(N,M|\{v\},\{u\})}=\frac{(-1)^{N-1}}{%
(u_N)^{N-1}}\left\{ \prod_{t<N}\frac{(u_N)^2-(u_t)^2}{(u_t)^2}\right\} \frac{%
\det Q}{\det H}.  \label{mcaf}
\end{equation}
The entries of $N\times N$ matrix $Q$ are given by
\begin{equation}
Q_{jN} =(v_j)^{M+N-1-2m}, \,\,\,\
Q_{jk} =H_{jk},\,\,k\neq N,
\end{equation}
where $H_{jk}$ are the matrix elements (\ref{t}). After the parametrization (%
\ref{hlip}) we find that the probability of the height $\pi _{1N}$
to be equal to $m$ is
\begin{eqnarray}
\langle m\rangle _{1N} &=&\frac{P_{1N}(m|\{q\})}{S(N,M|\{q\})}
\label{mcafh}\\
\ &=&q^{-\frac{(N-1)^2}2}\left\{ \prod_{t<N}\left(
1-q^{N-t-2}\right) \right\} \frac{\det {\cal Q}}{\det {\cal H}},
\nonumber
\end{eqnarray}
where
\begin{equation}
{\cal Q}_{jN} =q^{-\frac{j(M+N-1-2m)}2}, \,\,\,\,
{\cal Q}_{jk} ={\cal H}_{jk},\,\,k\neq N,
\end{equation}
and ${\cal H}_{jk}$ are the matrix elements (\ref{hme}). It is
evident that this expectation value is the same for the height
$\pi _{N1}$ in the opposite corner of the diagram.

The correlation function of the heights of the columns $\pi _{1N},\pi _{N1}$ at the opposite sides of the Young
diagram (see FIG. 9) may be obtained from the following scalar product
\begin{equation}
P_{N1;1N}(n;m|\{v\},\{u\})=\langle 0|\phi
_n\prod_{j=1}^{N-1}C(v_j)\prod_{j=1}^{N-1}B(u_j)\phi _m^{\dagger
}|0\rangle . \label{mncf}
\end{equation}
The scalar products (\ref{decsc}) and (\ref{mncf}) may be
expressed in terms of the skew Schur functions (\ref{ssf}) as
well.

The other function of interest is the projection of the Bethe wave function $%
\prod_{j=1}^NB(u_j)|0\rangle $ on the ''steady state'' vector:
\begin{equation}
|P\rangle =\sum_{{0\leq n_0,n_1,...,n_M\leq N \atop n_0+n_1+...+n_M=N}}\prod_{j=0}^M|n_j\rangle _j.  \label{stst}
\end{equation}
From the representation (\ref{cbawf}) and relation (\ref{cbafs})
we obtain the equality
\begin{equation}
\langle P|\prod_{j=1}^NB(u_j)|0\rangle =\left( u_1u_2\cdot
...\cdot u_N\right) ^{-M}\sum_{\lambda \subseteq
\{M^N\}}S_{\{\lambda \}}(u_1^2,u_2^2,...,u_N^2),  \label{ststbv}
\end{equation}
where the sum is over all partitions, $\lambda $ , into at most
$N$ parts each of which is less than or equal to $M$. It is known
that \cite{bres}
\begin{eqnarray}
\sum_{\lambda \subseteq \{M^N\}}S_{\{\lambda \}}(x_1,x_2,...,x_N) &=&\frac{%
\det \left( x_i^{j-1}-x_i^{M+2N-j}\right) }{\det \left(
x_i^{j-1}-x_i^{2N-j}\right) }  \label{bress} \\
&=&\prod_{i=1}^N\frac 1{1-x_i}\prod_{1\leq i<j\leq N}\frac
1{1-x_ix_j}, \nonumber
\end{eqnarray}
and we obtain for the projection (\ref{ststbv})
\begin{equation}
\langle P|\prod_{j=1}^NB(u_j)|0\rangle =\prod_{i=1}^N\frac 1{1-u_i^2}%
\prod_{1\leq i<j\leq N}\frac 1{1-u_i^2u_j^2}.  \label{ststo}
\end{equation}
By the construction the considered correlation function is the
generating function of the symmetric plane partitions, the plane
partitions satisfying
the condition $\pi _{ij}=\pi _{ji}$. From (\ref{ampl}) and the relation $%
t_k^d-t_k^a=l_k^d-l_k^a$ it follows that
\[
\langle P|\prod_{j=1}^NB(u_j)|0\rangle =\sum_{{\cal B}%
}u_1^{l_1^d-l_1^a}u_2^{l_2^d-l_2^a}\cdot ...\cdot u_N^{l_N^d-l_N^a}.
\]
The volume of the symmetric plane partitions may be expressed as
\begin{equation}
2|\pi |_{sym}=N^2M+\sum_{j=1}^N(2j-1)\left( l_j^d-l_j^a\right) ,
\label{volsym}
\end{equation}
where the sum is over the columns going along the ''$r$'' side of
the hexagon. Then
\[
\langle P|\prod_{j=1}^NB(q^{\frac{2j-1}2})|0\rangle =q^{-\frac{N^2M}2%
}\sum_{sym}q^{|\pi |},
\]
and we obtain the well known result for the generating function of
the symmetric plane partitions
\[
\sum_{\lambda \subseteq \{M^N\}}S_{\{\lambda
\}}(q,q^3,...,q^{2N-1})=\sum_{sym}q^{|\pi |}.
\]

Till now we have considered plane partitions in a box with $r=s$.
To study the general case when $r\neq s$ we have to consider the
following scalar products
\begin{eqnarray}
S^A(N,L,M|\{v\},\{u\},\{u^A\}) &=&\langle 0|\prod_{j=1}^NC(v_j)\prod_{j=1}^LA(u_j^A)\prod_{j=1}^NB(u_j)|0\rangle,
\label{cab} \\
S^D(N,L,M|\{v\},\{u\},\{u^D\}) &=&\langle 0|\prod_{j=1}^NC(v_j)\prod_{j=1}^LD(u_j^D)\prod_{j=1}^NB(u_j)|0\rangle.
\label{cdb}
\end{eqnarray}
Following the mapping introduced in this Section it may be shown
that these averages are the generating functions of plane
partitions in a box $r\times
s\times t$ with $r=N,s=N+L,t=M$ for (\ref{cab}), and $r=N+L,s=N,t=M$ for (%
\ref{cdb}).

\section{Boxed plane partitions and Toda lattice}

The $N\times N$ matrix ${\cal H}$ (\ref{hme}) is a H\"ankel matrix
with the matrix elements
\begin{equation}
{\cal H}_{jk}=h(q^{j+k-1}),  \label{hank}
\end{equation}
where
\begin{equation}
h(q)=\frac{q^{\frac{M+N}2}-q^{-\frac{M+N}2}}{q^{\frac
12}-q^{-\frac 12}}. \label{hqhq}
\end{equation}
By the successive subtraction of columns the determinant of this
matrix may be brought into the form
\begin{eqnarray}
\det {\cal H} &=&\det h, \label{hqh}\\
h_{jk} &=&D_q^{j+k-2}h(q),  \nonumber
\end{eqnarray}
and $D_q$ is the $q$-difference operator:
\begin{equation}
D_qf(z)=f(qz)-f(z).  \label{qd}
\end{equation}
Following the standard procedure \cite{ick}, \cite{ikk} it may be
shown that the function $\tau (N,M;q)\equiv \det {\cal H}$
satisfies the equation
\begin{equation}
D_q^2\ln \tau (N,M;q)=\frac{\tau (N+1,M;q)\tau (N-1,M;q)}{\tau ^2(N,M;q)}, \label{qeq}
\end{equation}
which, after the substitution
\[
\rho (N;q)=\ln \frac{\tau (N+1,M;q)}{\tau (N,M;q)}
\]
becomes the $q$-difference Toda equation:
\begin{equation}
D_q^2\rho (N;q)=e^{\rho (N+1;q)-\rho (N;q)}-e^{\rho (N;q)-\rho
(N-1;q)}. \label{qto}
\end{equation}
The role of time plays the deformation parameter $q$.

\section{Acknowledgments}
I would like to thank N.Yu. Reshetikhin, A.M. Vershik, and A.G.
Pronko for valuable discussions. This work was partially supported
by CRDF grant RUM1-2622-ST-04 and RFBR project 04-01-00825.

\end{document}